\documentclass[11pt]{article}      
\pdfoutput=1
\usepackage{amssymb}
\usepackage{amsmath}
\usepackage{amsthm}
\usepackage[]{graphicx}
\usepackage[]{subfigure}
\usepackage{tensor}
\usepackage{float}
\usepackage{color}
\usepackage{cancel}
\usepackage{setspace}
\usepackage{fancyhdr}
\usepackage{framed}
\usepackage{braket}
\usepackage{tikz}
\usepackage{comment}
\usepackage{bbold}
\usepackage{bbm}
\usepackage[noadjust]{cite}

\newcommand{\m}{\mathcal}

\usepackage[textwidth=6.5in,top=1.2in,bottom=1.2in]{geometry}

\usepackage{setspace}
\usepackage{simplewick}
\usepackage{float}
\usepackage{mathtools}
\usepackage{blkarray, bigstrut} %
\usepackage[bookmarks,linktocpage, colorlinks=true, plainpages = false, citecolor = red,  linkcolor=blue, urlcolor = magenta, filecolor = blue]{hyperref} 
\usepackage{braket}
\usepackage{mathrsfs}
\usepackage[title]{appendix}
\numberwithin{equation}{section}

\begin{document}

\thispagestyle{empty}
\begin{titlepage}
\nopagebreak

\title{\begin{center}\bf Spin-1/2 particles under the influence of a uniform magnetic field in the interior Schwarzschild solution\end{center}}

\vfill
\author{F.\,Hammad$^{1,2}$\footnote{\href{mailto:fhammad@ubishops.ca}{fhammad@ubishops.ca}},\, A.\,Landry$^3$\footnote{\href{mailto:alexandre.landry.1@umontreal.ca}{alexandre.landry.1@umontreal.ca}},\, P.\,Sadeghi$^1$\footnote{\href{mailto:psadeghi20@ubishops.ca}{psadeghi20@ubishops.ca}}}
\date{ }
\maketitle

\begin{center}
	\vspace{-0.7cm}
{\it  $\,^1$Department of Physics and Astronomy, Bishop's University, 2600 College Street, Sherbrooke, QC, J1M~1Z7 Canada}\\
	{\it  $\,^2$Physics Department, Champlain 
College-Lennoxville, 2580 College Street, Sherbrooke,  
QC, J1M~2K3 Canada}\\
{\it  $\,^3$D\'epartement de Physique, Universit\'e de Montr\'eal,\\
2900 Boulevard \'Edouard-Montpetit,
Montr\'eal, QC, H3T 1J4 Canada}
\end{center}
\bigskip

\begin{abstract}
The relativistic wave equation for spin-1/2 particles in the interior Schwarzschild solution in the presence of a uniform magnetic field is obtained. The fully relativistic regime is considered, and the energy levels occupied by the particles are derived as functions of the magnetic field, the radius of the massive sphere and the total mass of the latter. As no assumption is made on the relative strengths of the particles' interaction with the gravitational and magnetic fields, the relevance of our results to the physics of the interior of neutron stars, where both the gravitational and the magnetic fields are very intense, is discussed.
\end{abstract}

\end{titlepage}

\setcounter{page}{2}


\section{Introduction}\label{sec:I}
The behavior of charged quantum particles inside a uniform magnetic field in \linebreak \mbox{Minkowski spacetime} displays quantized energy levels ---\,the so-called Landau levels\,--- that were first derived by Rabi based on the relativistic Dirac equation \cite{Rabi} and by Landau based on the Schr\"odinger equation \cite{Landau} (see e.g., Ref.\,\cite{LandauLifshitz} for a textbook introduction). This discovery led to many physical applications in condensed matter physics, such as the explanation of the Landau diamagnetism \cite{Landau}, of the Shubnikov–de Haas effect \cite{Hadju}, of the de Haas-van Alphen effect \cite{Shoenberg} and of the integer quantum Hall effect \cite{QHEBook}. Remarkably, the Landau levels have even found astrophysical applications. In fact, the physics of neutron stars and other highly magnetized stellar objects is expected to be governed to some extent by the discrete nature of the energy levels occupied by the charged fermions moving under the influence of the intense magnetic field of such astrophysical objects \cite{BookNeutronStar,WhiteDwarfs,WhiteDwarfs3,Mass-Radius,WhiteDwarfs4,Broderick,Chamel1,Chamel2,Yakovlev, NeutronStarMatter}. When we recall that such highly compact astrophysical objects are also sources of very intense gravitational fields, it becomes extremely important to investigate the fate of Landau levels under the influence of a combination of a magnetic field and the gravitational field of a spherical object. Such a study has recently been conducted in Refs.\,\cite{GravityLandauI,GravityLandauII,QHGravity} based on the Klein-Gordon and Schr\"odinger equations by considering the gravitational interaction of the particle with a massive sphere as a small perturbation compared to the interaction of the particle with the magnetic field. Although relying on the Schr\"odinger equation and treating gravity perturbatively as done in these studies might be sufficient for laboratory \mbox{applications \cite{Landry1,Landry2,COWBall,Josephson},} when seeking astrophysical applications, implementing a general-relativistic and a non-perturbative approach and appealing to the relativistic Dirac equation are highly recommended.  
 
The Dirac equation in a curved spacetime caused {\it purely} by an intense magnetic field--known as the Melvin universe \cite{Melvin}---has been solved exactly in Ref.\,\cite{Nunes} where the quantized energy levels were extracted. Contemplating curved spacetimes caused purely by a magnetic field is motivated by the well-known fact that the strong magnetic fields of neutron stars could reach up to  {$10^{12-15}$}\,G  
 on the surface of magnetars \cite{Magnetars}, and up to $10^{17}\,$G in their interior (see, e.g., Refs.\,\cite{InsideNS1,InsideNS2} and the references therein). On the other hand, neutron stars are made of neutrons, as well as of a plasma of electrons, protons and muons, the energies of which are affected by such strong magnetic fields. The aim of the present paper is therefore to include the general-relativistic effect of the intense gravitational field of such stellar objects by studying the dynamics of charged fermions moving {\it inside} a massive spherical object in the presence of a uniform magnetic field. To the best of our knowledge, such a study has not appeared in the literature before. Unlike the purely magnetic spacetime considered in Ref.\,\cite{Nunes}, however, we consider in the present work the curved spacetime caused purely by the distribution of matter inside the core of the stellar objects rather than by the uniform magnetic field. The contribution of the latter to spacetime curvature will thus be considered negligible. The justification behind such an assumption will be discussed in Section \ref{sec:II}. 

The remainder of this paper is structured as follows. In Section \ref{sec:II}, we derive the Dirac equation in a general static and spherically symmetric curved spacetime in the presence of a uniform magnetic field. We discuss the various symmetries of the equation and the complications one faces when attempting to solve the equation exactly. In \mbox{Section \ref{sec:III}}, we apply the equation obtained in Section \ref{sec:II} to the case of the interior Schwarzschild solution. We then extract the quantized energy levels of charged fermions moving along the equator inside a massive spherical object by keeping only the leading terms of the differential equation. In Section \ref{sec:IV}, we briefly discuss the ways to overcome the limitations we encountered in solving our equation in Section \ref{sec:III}. In Section \ref{sec:V}, we apply our results to the study of the magnetization of neutron stars. We conclude this paper with a short discussion section in which we summarize our findings.

\section{Curved-Spacetime Dirac Equation in a Uniform Magnetic Field}\label{sec:II}
Since our aim in this paper is to consider the Dirac equation in the static and spherically symmetric interior Schwarzschild solution, let us start by deriving in this section the equation for a general static and spherically symmetric curved spacetime in the presence of a uniform magnetic field. For that purpose, let us write our spacetime metric in the following general form:
\begin{equation}\label{GeneralMetric}
    {\rm d}s^2=-\mathcal{A}^2(r)\,c^2{\rm d}t^2+\mathcal{B}^2(r)\,{\rm d}r^2+r^2({\rm d}\theta^2+\sin^2\theta{\rm d}\phi^2),
\end{equation}
where the radial functions $\mathcal{A}(r)$ and $\mathcal{B}(r)$ are everywhere regular except, maybe, at the origin $r=0$.  In addition, let us assume the constant and uniform magnetic field $\bf B$
to be parallel to the $z$-direction. A convenient gauge for the vector potential would then be $A_{\mu}=(0,0,0,\frac{1}{2}{Br^2\sin^2\theta})$. The Dirac equation of a particle of mass $m$ and of charge $e$ (that we take to be the negative charge of the electron for definiteness) moving inside a curved spacetime and minimally coupled to a Maxwell field $A_\mu$ reads \cite{Pollock,BookDiracInCS},
\begin{equation}\label{GeneralDirac}
i\hbar e^\mu_a\gamma^a\left(\partial_\mu+\frac{1}{8}\omega_\mu^{\;bc}\,[\gamma_b,\gamma_c]-\frac{ieA_\mu}{\hbar }\right)\psi=mc\psi.
\end{equation}
The quantities $e^\mu_a$ are the inverse of the spacetime tetrads $e^a_\mu$, whereas $\gamma^a$ are the gamma matrices, $\omega_\mu^{\,bc}$ is the spin connection and $[\gamma_b,\gamma_c]$ is the commutator of the gamma matrices. The spin connection is antisymmetric, $\omega_\mu^{\;bc}=-\omega_\mu^{\;cb}$, and it is expressed in terms of the tetrad fields and the Christoffel symbols $\Gamma_{\mu\nu}^\lambda$ of the spacetime by $\omega_\mu^{\;bc}=e^b_\nu\partial_\mu e^{c\nu}+\Gamma_{\mu\nu}^\lambda e^b_\lambda e^{c\nu}$.

\newpage
On the other hand, two choices for the four curved-spacetime gamma matrices $\gamma^\mu=e^a_\mu\gamma^a$ are possible. One may choose either to work with the tetrads' axes parallel to the $r$, $\theta$ and $\phi$ coordinates, or choose to orient the tetrads' axes parallel to some rectangular coordinate system \cite{BrillWheeler}. When combined with the standard representation of the constant flat-spacetime gamma matrices $\gamma^a$, given explicitly by
\begin{equation}\label{DiracMatrices}
\gamma^0=
\begin{pmatrix}
\mathbbm{1} & 0\\
0 & -\mathbbm{1}
\end{pmatrix},\qquad
\gamma^k=
\begin{pmatrix}
0 & \sigma_k\\
-\sigma_k & 0
\end{pmatrix},
\end{equation}
where $\mathbbm{1}$ is the $2\times2$ unit matrix and $\sigma_k$ (for $k=1,2,3$) are the three Pauli matrices, the curved-space gamma matrices $\gamma^\mu$ take on a simpler form in the first case than in the second \cite{BrillWheeler}. The spinor wavefunction $\psi$ in the two cases will be different, but physical quantities as well as the radial wave equation will all be the same as the two representations are simply related by a unitary transformation \cite{BrillWheeler}. However, to deal with the mixed cylindrical and spherical symmetries imposed, respectively, by the magnetic and gravitational fields, it is easier to adopt the spherical-coordinates representation of the gamma matrices, in which the three Pauli matrices take the form
\begin{equation}\label{PauliM}
\sigma_r=
\begin{pmatrix}
\cos\theta & \sin\theta e^{-i\phi}\\
\sin\theta e^{i\phi} & -\cos\theta
\end{pmatrix},\qquad
\sigma_\theta=
\begin{pmatrix}
-\sin\theta & \cos\theta e^{-i\phi}\\
\cos\theta e^{i\phi} & \sin\theta
\end{pmatrix},\qquad 
\sigma_\phi=
\begin{pmatrix}
0 & -ie^{-i\phi}\\
ie^{i\phi} & 0
\end{pmatrix}.
\end{equation}
We now choose the spacetime tetrads for the metric (\ref{GeneralMetric}) to be $e^a_\mu={\rm diag}(\mathcal{A},\mathcal{B},r,r\sin\theta)$. With these tetrads, we easily evaluate the nonvanishing components of the contracted spin connection with the commutator of the gamma matrices to be:
\begin{align}
    \tfrac{1}{8}\omega_0^{ab}[\gamma_a,\gamma_b]&=\frac{\mathcal{A}'}{2\mathcal{B}}\gamma^0\gamma^1,\qquad \tfrac{1}{8}\omega_2^{ab}[\gamma_a,\gamma_b]=\frac{1}{2\mathcal{B}}\gamma^1\gamma^2,\nonumber\\ \tfrac{1}{8}\omega_3^{ab}[\gamma_a,\gamma_b]&=\frac{\sin\theta}{2\mathcal{B}}\gamma^1\gamma^3+\frac{\cos\theta}{2}\gamma^2\gamma^3.
\end{align}
Plugging these, together with the vector potential $A_\mu$, into Eq.\,(\ref{GeneralDirac}) and then multiplying the whole equation on the left by $\beta=\gamma^0$ and using the definition $\alpha^k=\beta\gamma^k$ of the Dirac alpha matrices, the resulting equation reads,
\begin{equation}\label{ExplicitGeneralDirac}
      \left[\frac{i\partial_t}{\m{A}c}\psi+\frac{i\alpha^1}{\mathcal{B}}\left(\partial_r+\frac{1}{r}+\frac{\mathcal{A}'}{2\mathcal{A}}\right)+\frac{i\alpha^2}{r}\left(\partial_\theta+\frac{\cot\theta}{2}\right)+\frac{i\alpha^3}{r\sin\theta}\left(\partial_\phi-\frac{ieA_\phi}{\hbar}\right)\right]\psi=\frac{mc}{\hbar}\beta\psi.
\end{equation}
We are interested in this paper in stationary states of energy $E$, for which we have $\partial_t\psi=-\tfrac{i}{\hbar}E\psi$. On the other hand, we may simplify further Eq.\,(\ref{ExplicitGeneralDirac}) by setting $\psi=\Psi/(r\sqrt{\mathcal{A}\sin\theta})$. With these two replacements, the resulting equation in $\Psi$ takes the form
\begin{equation}\label{SimplifiedGeneralDirac}
    \Bigg[\frac{E}{\m{A}\hbar c}+\frac{i\alpha^1}{\mathcal{B}}\partial_r+\frac{i\alpha^2}{r}\partial_\theta+\frac{i\alpha^3}{r\sin\theta}\left(\partial_\phi-\frac{ieB}{2\hbar }r^2\sin^2\theta\right)-\frac{mc}{\hbar}\beta\Bigg]\Psi=0.\\
\end{equation}

This is the general equation obeyed by fermions under the influence of the uniform magnetic field. Unlike the well-known case of the Dirac equation in a spherically symmetric potential, i.e., in a central potential (see, e.g., Ref.\,\cite{Greiner}), our Eq.\,(\ref{SimplifiedGeneralDirac}) may not be separated in the radial variables $r$ and the angular variables $\theta$ and $\phi$. This is due to the mixture of the spherical symmetry imposed by the gravitational field and the cylindrical symmetry imposed by the vertical uniform magnetic field $\bf B$.
One might then be tempted to switch to cylindrical coordinates $(\rho,\phi,z)$ to benefit from the cylindrical symmetry pertaining to the magnetic field. Unfortunately, even doing so would not make Eq.\,(\ref{SimplifiedGeneralDirac}) easier to solve. In fact, by making the substitutions $\rho=r\sin\theta$ and $z=r\cos\theta$, Eq.\,(\ref{SimplifiedGeneralDirac}) would contain operators of the form $(\rho^2+z^2)^{-\frac{1}{2}}z\,\partial_\rho$ and $(\rho^2+z^2)^{-\frac{1}{2}}\rho\,\partial_z$. It is then evident that even if we neglect the $z$-variation of $\Psi$ by assuming $\partial_z\Psi\approx0$, i.e., by neglecting the vertical momentum of the particle, we would still be left with an equation that is not separable in the variables $\rho$, $z$ and $\phi$. 

It seems then that the only choice left is to set $z=0$ as well. Physically, this is due (in the {\it absence} of any other interaction of the particle) to the unstable nature of any position $z\neq0$, caused by the transverse gravitational attraction that pulls the particle away from any plane other than the equator. This observation can also be reached by 
referring to the variable $\theta$ in Eq.\,(\ref{SimplifiedGeneralDirac}). In fact, the only way of removing the $\theta$-dependence in that equation is to set $\theta=\frac{\pi}{2}$, i.e., to restrict ourselves to particles moving along the equatorial plane. This is because setting $\theta=\theta_0\neq
\frac{\pi}{2}$ would prevent us from letting $\rho=r\sin\theta_0$ vary and have at the same time planar motion. If $\theta_0$ is set to a constant other than $\frac{\pi}{2}$, then for $\rho$ to change, $r$ must change too, which makes the particle leave the plane $\theta=\theta_0$. Only along the equatorial plane $\theta_0=\frac{\pi}{2}$ does $\rho=r\sin\frac{\pi}{2}=r$ become a variable describing \mbox{planar motion}. 

We will come back to the possibility of considering a plane other than the equator in Section \ref{sec:IV}. Before setting $\theta=\frac{\pi}{2}$, however, we should first extract the general single differential equation from the coupled Dirac spinors.

\section{The Quantized Energy Levels}\label{sec:III}
To extract a quantization condition, if any, on the energy $E$ from Eq.\,(\ref{SimplifiedGeneralDirac}), we need first to decompose the Dirac spinor into a pair of $2$-spinors as follows
\begin{equation}
    \Psi(r,\theta,\phi)=
    \begin{bmatrix}
    \Phi(r,\theta,\phi)\\
    \Theta(r,\theta,\phi)
\end{bmatrix}.
\end{equation}
Substituting this ansatz into Eq.\,(\ref{SimplifiedGeneralDirac}), the latter splits into the following coupled first-order differential equations:
\begin{align}\label{SplitEqsSpinors}
\left(\frac{E}{\m{A}\hbar c}-\frac{mc}{\hbar}\right)\Phi+\left[\frac{i\sigma_r}{\mathcal{B}}\partial_r+\frac{i\sigma_\theta}{r}\partial_\theta+\frac{i\sigma_\phi}{r\sin\theta}\left(\partial_\phi-\frac{ieB}{2\hbar }r^2\sin^2\theta\right)\right]\Theta=0,\nonumber\\
\left(\frac{E}{\m{A}\hbar c}+\frac{mc}{\hbar}\right)\Theta+\left[\frac{i\sigma_r}{\mathcal{B}}\partial_r+\frac{i\sigma_\theta}{r}\partial_\theta+\frac{i\sigma_\phi}{r\sin\theta}\left(\partial_\phi-\frac{ieB}{2\hbar }r^2\sin^2\theta\right)\right]\Phi=0.
\end{align}
It is clear from these coupled differential equations that given the $\rho$-dependence of all the coupling terms, extracting one of the 2-spinors from one equation and substituting it into the other would result in a very complicated single equation. This is unlike what happens with the standard Dirac equation for the case of the central potential of the hydrogen atom \cite{Greiner}. It would be natural then at this stage to attempt to solve this system of equations by applying the approach developed in Refs.\,\cite{Alhaidari1,Alhaidari2,Alhaidari3}, which consists of applying a gauge transformation followed by a unitary transformation to turn one of the coupling terms in one of the two equations into a constant. Indeed, such a method considerably simplifies the resulting final single equation one arrives at. Unfortunately, such an approach cannot be consistently applied in our case because that approach is specifically designed only for a certain type of interactions of the particle with the gauge field, which is not the case for our present physical system\footnote{FH is grateful to Prof. A.\,D. Alhaidari for the helpful discussion on this subtle point.}. The other possibility would then be to attempt to rather adopt the approach developed in Ref.\,\cite{Alhaidari4} for the Dirac equation in curved spacetime. Unfortunately, that approach is not helpful to us either as it becomes rapidly involved even when the particle is not coupled to a gauge potential. For this reason, such an approach has been applied in Ref.\,\cite{Alhaidari4} only in $1+1$ dimensions. We are thus left only with the old strategy for dealing with coupled differential equations of the type (\ref{SplitEqsSpinors}).

Extracting the 2-spinor $\Theta$ from the second line of Eq.\,(\ref{SplitEqsSpinors}) and plugging it into the first, we obtain the following single equation for the 2-spinor $\Phi$:
\begin{align}\label{2-SpinorEq}
&\frac{\Phi_{,rr}}{\m{B}^2}+\left(\frac{2}{\m{B}r}-\frac{\m{B}_{,r}}{\m{B}^3}+\frac{\m{A}_{,r}E}{\m{A}\m{B}^2\left[E+\m{A}mc^2\right]}\right)\Phi_{,r}\nonumber\\
&+\frac{\Phi_{,\theta\theta}}{r^2}+\left(\frac{i\left[\m{B}-1\right]\sigma_\phi}{\m{B}r^2}+\frac{\cot\theta}{r^2}+\frac{i\sigma_\phi}{r}\frac{\m{A}_{,r}E}{\m{A}\m{B}\left[E+\m{A}mc^2\right]}\right)\Phi_{,\theta}\nonumber\\
&+\frac{\Phi_{,\phi\phi}}{r^2\sin^2\theta}
+\left(\frac{i\left[1-\m{B}\right]\sigma_\theta}{\m{B}r^2\sin\theta}-\frac{ieB}{\hbar}-\frac{i\sigma_\theta}{r\sin\theta}\frac{\m{A}_{,r}E}{\m{A}\m{B}\left[E+\m{A}mc^2\right]}\right)\Phi_{,\phi}\nonumber\\
&+\Bigg(\frac{E^2}{\m{A}^2\hbar^2c^2}-\frac{m^2c^2}{\hbar^2}-\frac{e^2B^2r^2\sin^2\theta}{4\hbar^2}-\frac{\left[\m{B}+1\right]eB\sigma_\theta\sin\theta}{2\hbar\m{B}}+\frac{\sigma_reB\cos\theta}{\hbar}\nonumber\\
&\qquad-\frac{\m{A}_{,r}EeBr\,\sigma_\theta\sin\theta}{2\hbar\m{A}\m{B}\left[E+\m{A}mc^2\right]}\Bigg)\Phi=0.
\end{align}
With this result, we can now proceed to restrict the motion of the particle to the equatorial plane by setting $\theta=\frac{\pi}{2}$. Therefore, we also need to perform the replacement $r\sin\theta=r\equiv\rho$ everywhere in this equation. On the other hand, for the $\theta$-derivatives, we use the following identities valid for motion restricted to a plane: $\partial_\theta=r\cos\theta\partial_\rho$ and $\partial_\theta^2=-r\sin\theta\partial_\rho+r^2\cos^2\theta\partial_\rho^2$. Therefore, for $\theta=\frac{\pi}{2}$ we should discard the partial derivative $\partial_\theta$ and perform the replacement  $\partial_\theta^2\rightarrow-\rho\,\partial_\rho$. The above equation then becomes
\begin{align}\label{Plane2-SpinorEq}
&\frac{\Phi_{,\rho\rho}}{\m{B}^2}+\left(\frac{2-\m{B}}{\m{B}\rho}-\frac{\m{B}_{,\rho}}{\m{B}^3}+\frac{\m{A}_{,\rho}E}{\m{A}\m{B}^2\left[E+\m{A}mc^2\right]}\right)\Phi_{,\rho}\nonumber\\
&+\frac{\Phi_{,\phi\phi}}{\rho^2}
+\left(\frac{i\left[\m{B}-1\right]\sigma_z}{\m{B}\rho^2}-\frac{ieB}{\hbar}+\frac{i\sigma_z}{\rho}\frac{\m{A}_{,\rho}E}{\m{A}\m{B}\left[E+\m{A}mc^2\right]}\right)\Phi_{,\phi}\nonumber\\
&+\left(\frac{E^2}{\m{A}^2\hbar^2c^2}-\frac{m^2c^2}{\hbar^2}-\frac{e^2B^2\rho^2}{4\hbar^2}+\frac{\left[\m{B}+1\right]eB\sigma_z}{2\hbar\m{B}}+\frac{\m{A}_{,\rho}EeB\rho\,\sigma_z}{2\hbar\m{A}\m{B}\left[E+\m{A}mc^2\right]}\right)\Phi=0.
\end{align}
We have used here the fact that according to Eq.\,(\ref{PauliM}), the spherical-coordinates Pauli matrix $\sigma_\theta$ reduces to $-\sigma_z$ for $\theta=\frac{\pi}{2}$ (as expected), where $\sigma_z$ is the usual third Pauli matrix in Cartesian coordinates. 

On the other hand, according to expressions (\ref{PauliM}) of the Pauli matrices, we see in Eq.\,(\ref{Plane2-SpinorEq}) that all the terms are diagonal and that the equation does not involve the angular variable $\phi$ explicitly. Therefore, without loss of generality, we can write the two independent solutions $\Phi_+(\rho,\phi)$ and $\Phi_-(\rho,\phi)$ of the equation in a basis made of the spin-eigenstates along the $z$-direction with eigenvalues $s=\pm1$, respectively. We thus introduce arbitrary radial functions $f_+(\rho)$ and $f_-(\rho)$ such that
\begin{equation}\label{Sigma3Basis}
\Phi_+(\rho)=e^{i\ell\phi}
\begin{bmatrix}
f_+(\rho)\\0
\end{bmatrix},\qquad 
\Phi_-(\rho)=e^{i\ell\phi}
\begin{bmatrix}
0\\f_-(\rho)
\end{bmatrix}.
\end{equation}
Imposing the periodic condition along the equator, $\Phi(\rho,\phi+2\pi)=\Phi(\rho,\phi)$, we learn that the quantum number $\ell$ must be an integer number, which we take here to be nonnegative in accordance with the negative sign we chose for the charge $e$. Plugging these ansatzes into Eq.\,(\ref{Plane2-SpinorEq}) yields the following second-order differential equation in $\rho$:
\begin{align}\label{fEq}
&\frac{f_s''}{\m{B}^2}+\left(\frac{2-\m{B}}{\m{B}\rho}-\frac{\m{B}'}{\m{B}^3}+\frac{\m{A}'E}{\m{A}\m{B}^2\left[E+\m{A}mc^2\right]}\right)f'_s\nonumber\\
&+\Bigg(\frac{E^2}{\m{A}^2\hbar^2c^2}-\frac{m^2c^2}{\hbar^2}+\frac{eB\ell}{\hbar}-\frac{e^2B^2\rho^2}{4\hbar^2}-\frac{\m{B}\ell^2+\left[\m{B}-1\right]s\ell}{\m{B}\rho^2}+\frac{\left[\m{B}+1\right]eBs}{2\hbar\m{B}}\nonumber\\
&\qquad-\frac{\m{A}'Es\left[2\hbar\ell-eB\rho^2\right]}{2\hbar\rho\m{A}\m{B}\left[E+\m{A}mc^2\right]}\Bigg)f_s=0,
\end{align}
Here, and henceforth, a prime denotes a derivative with respect to the variable $\rho$, and the function $f_s(\rho)$ stands for the two cases $f_\pm(\rho)$ of $s=\pm1$, respectively. Next, we introduce the following ansatz: \begin{equation}\label{FirstAnsatz}
    f_s(\rho)=\left(\frac{\mathcal{B}}{\mathcal{A}}E+\mathcal{B}mc^2\right)^{\frac{1}{2}}F_s(\rho),
\end{equation}
for some radial function $F_s(\rho)$. Please note that only the convergence of the radial function $F_s(\rho)$ is required henceforth, for the argument of the square root in this ansatz converges for both large and small values of $\rho$ as we shall see below after introducing the explicit components of the spacetime metric. Plugging now this ansatz into Eq.\,(\ref{fEq}), the latter takes the following form:
\begin{multline}\label{SimplifiedfEq}
\!\!\!\frac{F_s''}{\m{B}^2}+\frac{2-\m{B}}{\m{B}\rho}F'_s+\bigg[\frac{E^2}{\m{A}^2\hbar^2c^2}-\frac{m^2c^2}{\hbar^2}+\frac{eB\ell}{\hbar}-\frac{e^2B^2\rho^2}{4\hbar^2}-\frac{\m{B}\ell^2+(\m{B}-1)s\ell}{\m{B}\rho^2}\\
+\frac{(\m{B}+1)eBs}{2\hbar\m{B}}-\frac{\m{A}'Es(2\hbar\ell-eB\rho^2)}{2\hbar\rho\m{A}\m{B}(E+\m{A}mc^2)}-\frac{1}{2\m{B}^2}\left(\frac{\m{A}'E}{\m{A}[E+\m{A}mc^2]}-\frac{\m{B}'}{\m{B}}\right)'\\
-\left(\frac{\m{A}'E}{2\m{A}\m{B}\left[E+\m{A}mc^2\right]}-\frac{\m{B}'}{2\m{B}^2}\right)^2-\frac{2-\m{B}}{2\m{B}\rho}\left(\frac{\m{A}'E}{\m{A}\left[E+\m{A}mc^2\right]}-\frac{\m{B}'}{\m{B}}\right)\bigg]F_s=0.
\end{multline}
Finally, the following ansatz,
\begin{equation}\label{FinalAnsatz}
    F_s(\rho)=G_s(\rho)\exp\left[\int\frac{(1-\m{B})^2}{2\rho}{\rm d}\rho\right],
\end{equation}
for an arbitrary radial function $G_s(\rho)$, allows us to convert Eq.\,(\ref{SimplifiedfEq}) into the following more useful form:
\begin{align}\label{FinalfEq}
&G''_s+\frac{G'_s}{\rho}+\bigg[
\frac{E^2}{\m{A}^2\hbar^2c^2}-\frac{m^2c^2}{\hbar^2}+\frac{eB\ell}{\hbar}-\frac{e^2B^2\rho^2}{4\hbar^2}-\frac{\ell^2}{\rho^2}+\frac{(1-\m{B})s\ell}{\m{B}\rho^2}+\frac{(1+\m{B})eBs}{2\hbar\m{B}}\nonumber\\
&-\frac{\m{A}'Es\left[2\hbar\ell-eB\rho^2\right]}{2\hbar\rho\m{A}\m{B}\left[E+\m{A}mc^2\right]}-\frac{1}{2\m{B}^2}\left(\frac{\m{A}'E}{\m{A}\left[E+\m{A}mc^2\right]}-\frac{\m{B}'}{\m{B}}\right)'-\left(\frac{\m{A}'E}{2\m{A}\m{B}\left[E+\m{A}mc^2\right]}-\frac{\m{B}'}{2\m{B}^2}\right)^2\nonumber\\
&-\frac{2-\m{B}}{2\m{B}\rho}\left(\frac{\m{A}'E}{\m{A}\left[E+\m{A}mc^2\right]}-\frac{\m{B}'}{\m{B}}\right)-\frac{\m{B}'(1-\m{B})}{\m{B}^2\rho}-\frac{(1-\m{B})^4}{4\m{B}^2\rho^2}\bigg]\m{B}^2G_s=0.
\end{align}

It is clear that Eq.\,(\ref{FinalfEq}) would not be easy to solve exactly. However, we can clearly see now which terms of the equation could safely be dropped out even without making any prior assumption on the orders of magnitude of the various physical quantities involved, such as the energy of the particle and the strengths of the gravitational and magnetic interaction terms. In Section \ref{sec:IV}, we shall come back to this point to discuss the order of magnitude of the correction that would have been brought to our final result for the energy levels of the particle if we had kept those extra terms.

Now, while Eq.\,(\ref{FinalfEq}) seems analytically challenging to solve even when keeping only the dominant terms, the fact that the metric components $\mathcal{A}^2$ and $\mathcal{B}^2$ of the interior Schwarzschild solution depend on $\rho^2$ instead of $\rho$ will greatly simplify our task as we shall see shortly. For a spherical body of mass $M$, of radius $R$ and of uniform density, the interior Schwarzschild solution describing the gravitational field inside the body is given by the metric (\ref{GeneralMetric}), with \cite{Synge}
\begin{equation}\label{InteriorSchwar}
    \mathcal{A}(\rho)=\frac{3}{2}\left(1-\frac{r_s}{R}\right)^{\frac{1}{2}}-\frac{1}{2}\left(1-\frac{r_s\rho^2}{R^3}\right)^{\frac{1}{2}},\qquad \mathcal{B}(\rho)=\left(1-\frac{r_s\rho^2}{R^3}\right)^{-1/2},
\end{equation}
where $r_s=2GM/c^2$ is the Schwarzschild radius of the massive body. For convenience, we introduce the dimensionless parameter $\eta=\frac{3}{2}(1-\frac{r_s}{R})^{\frac{1}{2}}$ and the inverse length squared $\lambda=\frac{r_s}{R^3}$. Please note that by choosing this metric, we have ignored the possible geometric effect of the magnetic field on the spacetime. We will come back to this point in Section \ref{sec:IV} as well.   

Let us now show the convergence of the square root in our ansatz (\ref{FirstAnsatz}). Please note that for very small values of $\rho$, the functions (\ref{InteriorSchwar}) approximate to $\mathcal{A}\sim\eta-\frac{1}{2}+\frac{\lambda}{4}\rho^2$ and $\mathcal{B}\sim1+\frac{\lambda}{2}\rho^2$, from which we see that the leading-order $\rho$-dependent term in the argument of the square root in Eq.\,(\ref{FirstAnsatz}) is proportional to $\rho^2$. This establishes thus the convergence of that square root for very small $\rho$. On the other hand, for $\rho\geq R$, the functions (\ref{InteriorSchwar}) lead, by continuity, to the metric components of the {\it exterior} Schwarzschild solution, for we have then $\mathcal{A}=\mathcal{B}^{-1}=(1-\frac{r_s}{\rho})^{\frac{1}{2}}$. Therefore, for very large values of $\rho$, the functions $\mathcal{A}$ and $\mathcal{B}$ in the argument of the square root in our ansatz (\ref{FirstAnsatz}) approximate to $\mathcal{A}\sim1-\frac{r_s}{2\rho}$ and $\mathcal{B}\sim1+\frac{r_s}{2\rho}$. This implies that the leading-order $\rho$-dependent term in the argument of the square root is proportional to $1/\rho$. This establishes thus the convergence of that square root for large values of $\rho$ as well, whence we deduce the convergence of that square root for all values of $\rho$. 

Next, we use the functions (\ref{InteriorSchwar}) to expand $1/\m{A}$ and $1/\mathcal{A}^2$, up to the first order in the parameter $\lambda$, as follows:
\begin{align}\label{ABPowerSeries}
    \frac{1}{\m{A}}&\sim\frac{2}{(2\eta-1)}-\frac{\lambda\rho^2}{(2\eta-1)^2},\nonumber\\
    \frac{1}{\m{A}^{2}}&\sim\frac{4}{(2\eta-1)^2}-\frac{4\lambda\rho^2}{(2\eta-1)^3}.
\end{align}
Expanding only up to the first order in $\lambda$ is justified by the order of magnitude of $\lambda$ (even for neutron stars as we will see in Section \ref{sec:IV}). On the other hand, we also have $1/\m{B}\sim1-\frac{\lambda}{2}\rho^2$ and $\m{B}^2\sim1+\lambda\rho^2$. With these expansions, Eq.\,(\ref{SimplifiedfEq}) takes, up to the first order in the parameter $\lambda$, the following form:
\begin{align}\label{1stOrderFinalfEq}
&\frac{{\rm d}^2G_s}{{\rm d}\rho^2}+\frac{1}{\rho}\frac{{\rm d}G_s}{{\rm d}\rho}+\Bigg[
\frac{4E^2}{(2\eta-1)^2\hbar^2c^2}-\frac{m^2c^2}{\hbar^2}+\frac{eB(\ell+s)}{\hbar}+\frac{\lambda eBs}{4\hbar}-\frac{\lambda(2\ell^2+s\ell-2)}{2}\nonumber\\
&-\frac{2\lambda(s\ell+1)}{(2\eta-1)[2+(2\eta-1)\frac{mc^2}{E}]}-\frac{\ell^2}{\rho^2}-\Bigg(\frac{e^2B^2}{4\hbar^2}+\frac{\lambda m^2c^2}{\hbar^2}+\frac{8\lambda E^2[\eta-1]}{\hbar^2c^2[2\eta-1]^3}+\frac{\lambda eB[2\ell+s]}{2\hbar}\nonumber\\
&+\frac{\lambda eBs}{\hbar(2\eta-1)[2+(2\eta-1)\frac{mc^2}{E}]}\Bigg)\rho^2-\frac{\lambda e^2B^2}{4\hbar^2}\rho^4\Bigg]G_s=0.
\end{align}
This equation has the form
\begin{equation}\label{GEquation}
    G_s''+\frac{G_s'}{\rho}+\left(\varepsilon-\frac{\ell^2}{\rho^2}-\kappa^2\rho^2-\tau^2\rho^4\right)G_s=0,
\end{equation}
where a prime denotes a derivative with respect to $\rho$ and the constants $\varepsilon$, $\kappa^2$ and $\tau^2$ are straightforwardly read off from Eq.\,(\ref{1stOrderFinalfEq}). Setting $G_s=\chi_s/\sqrt{\rho}$ in Eq.\,(\ref{GEquation}), the latter takes the form
\begin{equation}\label{Anharmonic}
    \chi''_s+\left(\varepsilon-\frac{\ell^2-\frac{1}{4}}{\rho^2}-\kappa^2\rho^2-\tau^2\rho^4\right)\chi_s=0.
\end{equation}
This is a Schr\"odinger equation with a centrifugal barrier term and an isotropic quartic anharmonic oscillator potential term. Please note that (i) in the case of a weak gravitational interaction compared to the magnetic one, i.e., for $\lambda\ll |e|B/\hbar$, and/or (ii) in the case of charged fermions moving very close to the center of the spherical mass, i.e., for $\rho\ll R$, the purely quartic term $\tau^2\rho^4$ can be dropped out. In that case, Eq.\,(\ref{Anharmonic}) reduces to the one that has already been solved exactly in Ref.\,\cite{GravityLandauI}, where the quantized eigenvalues $\varepsilon_{n,\ell}$ have been extracted. We will come back to these special cases in Section \ref{sec:IV-1}. Here, we will extract the eigenvalues from Eq.\,(\ref{Anharmonic}) by keeping the quartic term.

It is shown in Ref.\,\cite{JMP(1986)} that even though  Eq.\,(\ref{Anharmonic}) is not solvable exactly, one may still extract an approximate analytical expression for its eigenvalues using the Jeffreys–Wentzel–Kramers–\\Brillouin (JWKB) approximation. To the fourth order in the approximation, the result reads \cite{JMP(1986)}
\begin{equation}\label{QuantizedEpsilon}
\varepsilon_{n,\ell,s}=(2 \tau)^{\frac{2}{3}}\sum_{k=0}^8\left[\frac{3\sqrt{\pi}\,\Gamma(\tfrac{3}{4})}{\sqrt{2}\,\Gamma(\tfrac{1}{4})}\left(2n+1+\ell\right)\right]^{\frac{4-2k}{3}}N_k\left(\kappa,\tau,\ell\right),
\end{equation}
where the explicit expressions of the nine terms $N_k(\kappa, \tau,\ell)$ are given in Appendix \ref{sec:App}. It is clear from this expression that the usual flat-space Landau levels of a charged fermion moving under the influence of a uniform magnetic field are dramatically altered. In fact, instead of the splitting of the levels that occurs due to a weak gravitational field \mbox{outside \cite{GravityLandauI}} or inside a spherical mass \cite{QHGravity}, what happens here is a complete redistribution of the levels. Both the separation of the latter and their strengths are redefined for each principle quantum number $n$. 

For the sake of illustration and simplicity, we display here explicitly the expression we obtain when keeping only the first term of the sum in Eq.\,(\ref{QuantizedEpsilon}). By substituting the constants $\varepsilon$, $\kappa$ and $\tau$, as well as the term $N_0=1$ (as given in Appendix \ref{sec:App}) into \mbox{Eq.\,(\ref{QuantizedEpsilon}),} we find
\begin{align}\label{QEpsilonN0}
   E_{n,\ell,s}&=\left(\eta-\frac{1}{2}\right)\Bigg\{m^2c^4-\hbar c^2eB\left(\ell+s+\frac{\lambda s}{4}\right)+\frac{\lambda\hbar^2c^2(2\ell^2+s\ell-2)}{2}\nonumber\\
   &+\frac{2\lambda\hbar^2c^2(s\ell+1)}{(2\eta-1)[2+(2\eta-1)\frac{mc^2}{E}]}+\hbar^2c^2\left[\left(2n+1+\ell\right)\frac{3\,\Gamma(\tfrac{3}{4})}{\Gamma(\tfrac{1}{4})}\right]^{\frac{4}{3}}\left(\frac{\pi eB\sqrt{\lambda}}{2\hbar}\right)^\frac{2}{3}\Bigg\}^{\frac{1}{2}}.
\end{align}
This does not, in fact, look at all like the familiar Landau levels. It must be noted here that this formula does not reduce to the flat-space result when setting $\lambda=0$ because the expansion in terms of the JWKB integrals is about the pure quartic oscillator levels, not the other way around \cite{JMP(1986)}. Therefore, to achieve a high accuracy in the Formula (\ref{QuantizedEpsilon}), $\kappa^2$ should not be larger than $\tau^2$. To properly recover the flat-space case, one needs to set $\lambda=0$ in Eq.\,(\ref{1stOrderFinalfEq}). Similarly, when the gravitational interaction is weaker than the magnetic interaction, one needs to extract the corresponding Landau levels by starting from Eq.\,(\ref{1stOrderFinalfEq}) as we shall do now.
\subsection{Weak Quartic Term}\label{sec:IV-1}
When the gravitational interaction is weak compared to the magnetic interaction and/or when the charged particles move very closely to the center of the spherical mass, the quartic term $\tau^2\rho^4$ in Eq.\,(\ref{GEquation}) can safely be dropped out so that the equation takes the form,
\begin{equation}\label{GEquationWithoutQuartic}
    G_s''+\frac{G_s'}{\rho}+\left(\varepsilon-\frac{\ell^2}{\rho^2}-\kappa^2\rho^2\right)G_s=0.
\end{equation}
The solution to this equation is easily found to be given in terms of the confluent hypergeometric function as follows (see Ref.\,\cite{GravityLandauI} for the details of the derivation):
\begin{equation}
    G_s(\rho)=C\rho^{\ell}e^{-\frac{|\kappa|}{2}\rho^2}\,_1F_1\left(\frac{1+\ell}{2}-\frac{\varepsilon}{4|\kappa|};\ell+1;|\kappa|\rho^2\right).
\end{equation}
This solution is one of the two independent solutions of Eq.\,(\ref{1stOrderFinalfEq}) that is finite at the origin $\rho=0$. $C$ is one of the constants of integration, and$\,_1F_1(a;b;\xi)$ is the confluent hypergeometric function of a variable $\xi$ for any values of the parameters $a$ and $b$, except when $b$ is a negative integer for which case the function has a simple pole \cite{HandBook}. Therefore, to guarantee a square-integrable character for the wavefunction, the following condition must be imposed for an arbitrary positive integer $n$:
\begin{equation}
    \frac{1+\ell}{2}-\frac{\varepsilon}{4|\kappa|}=-n,
\end{equation}
for which case, the confluent hypergeometric function becomes indeed a polynomial of finite degree $n$. By substituting into this condition, the values of $\varepsilon$ and $\kappa$ as read off from \mbox{Eq.\,(\ref{1stOrderFinalfEq})}, we arrive at the following condition involving the energy $E$ of the \mbox{charged fermion:}
\begin{align}
&\frac{4E^2}{(2\eta-1)^2\hbar^2c^2}-\frac{m^2c^2}{\hbar^2}+\frac{eB(\ell+s)}{\hbar}+\frac{\lambda eBs}{4\hbar}-\frac{\lambda(2\ell^2+s\ell-2)}{2}\nonumber\\
&-\frac{2\lambda(s\ell+1)}{(2\eta-1)[2+(2\eta-1)\frac{mc^2}{E}]}=(2n+1+\ell)\Bigg[\frac{e^2B^2}{\hbar^2}+\frac{4\lambda m^2c^2}{\hbar^2}+\frac{32\lambda E^2(\eta-1)}{\hbar^2c^2(2\eta-1)^3}\nonumber\\
&+\frac{2\lambda eB(2\ell+s)}{\hbar}+\frac{4\lambda eBs}{\hbar(2\eta-1)(2+[2\eta-1]\frac{mc^2}{E})}\Bigg]^{\frac{1}{2}}.
\end{align}
We may solve this equation for $E$ by keeping again only the first-order terms in the parameter $\lambda$, to arrive at the following quantization condition on the energy $E$ of \mbox{the particle}:
\begin{align}\label{FinalQuantization}
E_{n,\ell,s}&=\hbar c\left(\eta-\frac{1}{2}\right)\Bigg\{\frac{m^2c^2}{\hbar^2}-\frac{eB(\ell+s)}{\hbar}-\frac{\lambda eBs}{4\hbar}+\frac{\lambda(2\ell^2+s\ell-2)}{2}\nonumber\\
&\quad+\frac{\lambda(s\ell+1)}{(2\eta-1)}\frac{\sqrt{1+(2n+1-s)\frac{\hbar eB}{m^2c^2}}}{1+\sqrt{1+(2n+1-s)\frac{\hbar eB}{m^2c^2}}}+(2n+1+\ell)\Bigg[\frac{e^2B^2}{\hbar^2}+\frac{4\lambda m^2c^2}{\hbar^2}\nonumber\\
&\quad+\frac{2\lambda eB(2\ell+s)}{\hbar}+(\eta-1)\frac{8\lambda(m^2c^2+[2n+1-s]\hbar eB)}{\hbar^2(2\eta-1)}\nonumber\\
&\quad+\frac{2\lambda eBs}{\hbar(2\eta-1)}\frac{\sqrt{1+(2n+1-s)\frac{\hbar eB}{m^2c^2}}}{1+\sqrt{1+(2n+1-s)\frac{\hbar eB}{m^2c^2}}}\Bigg]^{\frac{1}{2}}\Bigg\}^{\frac{1}{2}}.
\end{align}

We readily note that for $M=0$ (i.e., by setting $\eta=\frac{3}{2}$ and $\lambda=0$), the result (\ref{FinalQuantization}) reduces to the usual quantized energy levels of a charged fermion inside a uniform magnetic field in Minkowski spacetime: $E_{n,s}=[m^2c^4+\hbar c^2(2n+1-s\frac{e}{|e|})|e|B]^{\frac{1}{2}}$. In addition, we also note, by setting $B=0$, that even in the absence of the magnetic field the gravitational field induces quantized energy levels. Furthermore, we neatly recognize a general-relativistic correction to the pure relativistic Landau levels arising from the magnetic field. Such a correction arises from the factor $(\eta-\frac{1}{2})$ multiplying the right-hand side. 

We are now going to discuss the ways to overcome the limitations we encountered in this section and what modifications would be brought to the results we obtained so far.

\section{Refinements and Prospective Extensions}\label{sec:IV}
Two limitations, we could not avoid when deriving our results in the previous section, were (i) the fact that we considered only the possible motion of the particle along the equator inside the massive sphere and (ii) the fact that we ignored the geometric effect that might be caused to spacetime by the magnetic field. The two other limitations we imposed on our analysis were (iii) the uniform mass density we assumed for the interior of the spherical mass and (iv) the first-order approximation in $\lambda$ we limited ourselves to when writing Eq.\,(\ref{1stOrderFinalfEq}) to be able to solve the latter exactly. Our aim in this section is to address these limitations and argue that they do not actually alter much the actual physical conclusions we would have reached had we not adopted those various approximations.

The first limitation consists of the restricted number of degrees of freedom we imposed on the charged fermion. By restricting the motion of the particle to the equatorial plane inside the massive sphere, we implicitly assumed the particle's motion to be confined to such a plane due to the collective interaction of the particles among themselves. However, it is actually possible, for the same reason, to also consider the possibility of motion along a plane other than the equatorial plane. In fact, the vertical pull of gravity is then balanced by these other interactions so that only the horizontal variation of the gravitational potential matters. For such a case, however, we need to switch entirely to cylindrical coordinates. The interior Schwarzschild solution needs to be expressed in such coordinates and we need to set $r^2=\rho^2+z_0^2$ for some vertical distance $z_0$ from the equator in the resulting differential equation that would be similar in form to Eq.\,(\ref{2-SpinorEq}). As the differential operator $\partial_\theta$ would be replaced by $z_0\partial_\rho$ and the operator $\partial_\theta^2$ would be replaced by $z_0^2\partial_\rho^2-(\rho+z_0)\partial_\rho$, the equation would still be a second-order differential equation in the single variable $\rho$. A first-order approximation in the parameter $\lambda$ might then be adopted. Although such an equation would certainly not be as easy to solve as Eq.\,(\ref{1stOrderFinalfEq}), we expect the energy of the particle to be only shifted by a constant while obeying a quantization condition similar to our results (\ref{QuantizedEpsilon}) and (\ref{FinalQuantization}). A further refinement consists then of considering the case $\partial_z\Psi\neq0$ and a non-constant $z$. Unfortunately, the fact that two variables would then be involved, no simple procedure could be expected. A separate future work should specifically be devoted to solving that case using cylindrical coordinates. Indeed, the problem of solving our very general Eq.\,(\ref{2-SpinorEq}) without restricting the latter to the equatorial plane might be very important for other astrophysical applications besides neutron stars. We believe that the problem might very well be tackled by adapting to our case the existing numerical methods that have already proved fruitful elsewhere for solving such equations (see, e.g., the recent work \cite{Numerical} and the references therein).

The second interesting refinement that should be discussed here is the possibility of adding to the metric components, given by Eq.\,(\ref{InteriorSchwar}), the spacetime curvature caused by the magnetic field. The geometric effect of combining the external gravitational field sourced by a spherical mass with a uniform magnetic field $B$ gives rise to the so-called Schwarzschild–Melvin spacetime \cite{Ernst}. Therefore, an analogue modification to the interior Schwarzschild metric components given by Eq.\,(\ref{InteriorSchwar}) is also expected when including the spacetime curvature caused by $B$. However, since the corrections brought to the familiar exterior Schwarzschild solution by the magnetic field are of the order of $\epsilon_0GB^2/c^2$, where $\epsilon_0$ is the vacuum permittivity constant \cite{Galtsov}, we expect the correction that would be brought to the parameter $\lambda$ in our result (\ref{FinalQuantization}) to be of the order of $\sim$$\epsilon_0GB^2/c^2$ as well. For a neutron star of mass $\sim$$1.4\,M_\odot$ and of radius $10$\,km, the Schwarzschild radius is about $\sim$$4.1\,$km for which we find that $\lambda\sim0.004\,$km$^{-2}$. On the other hand, for such a neutron star with an interior magnetic field as strong as $\sim$$10^{17}\,$G, we have $\epsilon_0GB^2/c^2\sim7\times10^{-9}\,$km$^{-2}$. The expected correction is thus quite insignificant.

The third limitation arises from having discarded in Eq.\,(\ref{1stOrderFinalfEq}) terms of the second order and higher in the parameter $\lambda$. We can justify now such an approximation by the fact that, as we just saw for a typical neutron star, $\lambda\sim4\times10^{-3}\,$km$^{-2}$. This entails that if we had kept those higher order terms the expected correction to be brought to the quantized energy levels (\ref{FinalQuantization}) can be obtained using time-independent perturbation theory in the manner worked out in Refs. \,\cite{GravityLandauI,GravityLandauII} for the case of Landau levels in the exterior Schwazschild solution. 

Another refinement that should be taken into account is the fact that matter density inside neutron stars is not really uniform as we have assumed it in our calculations in the previous section. The interior matter density increases with increasing depth. However, the core of a neutron star makes up the largest part of the star and may be subdivided into outer and inner parts, each characterized by slowly varying mass densities with depth \cite{NSDensity}. For this reason, although considering a uniform mass density is only a simplifying model for the interior of a neutron star, we expect that the conclusions that would be derived by assuming a slowly varying $\rho$-dependent mass density would not differ much from those obtained here. Unfortunately, an analysis based on the interior Schwarzschild solution with a $\rho$-dependent mass density is beyond the scope of the present paper and should be considered in future works as well. It must also be noted here that numerical methods would certainly be beneficial for such a task.

The last two extensions of this work we would like to mention here is the possibility of studying the effect of the interior Schwarzschild solution on (i) charged fermions when the magnetic field is nonuniform and (ii) on the neutron star matter when the latter is taken to be a superfluid \cite{NSSuperfluid}. For the former case, one needs to use again numerical methods (as in Ref.\,\cite{NonUniformB}) to solve the resulting equation, whereas for the latter case, one needs to consider, instead of the Dirac equation, the Gross–Pitaevskii equation coupled to the magnetic field inside the interior Schwarzschild solution.     

\section{Applications}\label{sec:V}
Despite the limitations we discussed above, it is still interesting to apply our results to some systems that rely on the Landau energy levels of charged fermions. We will first examine the charged fermions inside a neutron star, and then we examine charged fermions inside a laboratory-made spherical mass.

\subsection{Inside a Neutron Star}
We propose to use here our results (\ref{QuantizedEpsilon}) and (\ref{FinalQuantization}) to examine the effect of gravity on the magnetization $\mathscr{M}$ inside a neutron star.  The magnetization $\mathscr{M}$ is given by the following general formula \cite{Broderick}:
\begin{equation}\label{GeneralM}
    \mathscr{M}=\sum_{i=e,\mu,p}\left(\frac{\partial \epsilon_i}{\partial B}-\mu^i_f\frac{\partial n_i}{\partial B}\right),
\end{equation}
where the summation is over the electrons, muons and protons, respectively; $\epsilon_i$ are the energy densities of those charged fermions and $n_i$ are their number densities. The magnetic field $B$ in this formula is the original seed field (magnetizing field) that is thus gradually modified by the magnetization into $H=B+4\pi\mathscr{M}$, in CGS units \cite{LLPTextbook}. A strong magnetization has been suggested by some authors to play a role even in giving rise to the original (primal) magnetic field itself \cite{Dong2013}. However, despite considerable research on the topic, there is in the literature no consensus yet on the origin of such strong seed magnetic fields at the surface nor in the core of such compact stars. For a quick guide through the various proposals put forward in the literature, see, e.g., the reviews \cite{ReviewEoS2015,Universe2021}. 

Considering the magnetizing field $B$ to be constant in time and uniform over the radius of the star, as we do in this paper, is also only an approximation. However, given that the relaxation times of the star's magnetic field decay is of the order of $\sim$$10^3$ \mbox{years \cite{Universe2018}}, our present assumption of a constant magnetic field fits amply within the early stages of the magnetic evolution of the star. Similarly, assuming a slow variation of the magnetic field with distance inside highly magnetized stars as presented, for example, in Ref.\,\cite{BDistribution} is still consistent with our present purpose of providing a preliminary study on the effect of gravity on the magnetic properties of such compact stellar objects.

First, recall that when ignoring the contribution of the gravitational field, the quantities $\epsilon_i$ and $n_i$ are given by \cite{Broderick}
\begin{align}\label{EnergyAndNumberDensity}
    \epsilon_i&=\frac{|e|B}{4\pi^2}\sum_s\sum_{n=0}^{n_{max}}\left(\mu^i_fk^i_{f,n,s}+\tilde{m}^{i2}_{n,s}\ln\left|\frac{\mu^i_f+k^i_{f,n,s}}{\tilde{m}^i_{n,s}}\right|\right),\nonumber\\
    n_i&=\frac{|e|B}{2\pi^2}\sum_s\sum_{n=0}^{n_{max}}k^i_{f,n,s},
\end{align}
where
\begin{align}\label{mTilde}
    \tilde{m}^{i2}_{n,s}&=m^{i2}+\left(2n+1-\frac{e}{|e|}s\right)|e|B,\nonumber\\
    k_{f,n,s}^{i2}&=\mu_f^{i2}-\tilde{m}_{n,s}^{i2}.
\end{align}
The Fermi energies $\mu^{i}_f$ are fixed by the chemical potentials and the maximum integer $n_{max}$ in the summations represents the integer preceding the value of $n$ for which $k_{f,n,s}^{i2}$ becomes negative. As the equations are more involved, we will work in this subsection in the natural units $\hbar=c=1$. Formula (\ref{GeneralM}) then gives \cite{Broderick} \begin{equation}\label{FlatMagnetization}
    \mathscr{M}=\sum_{i=e,\mu,p}\left[\frac{\epsilon_i-\mu^i_fn_i}{B}+\frac{B}{2\pi^2}\sum_{s}\sum_{n=0}^{n_{max}}\left(n+\frac{1}{2}-\frac{s}{2}\right)\ln\left|\frac{\mu^i_f+k^i_{f,n,s}}{\tilde{m}^i_{n,s}}\right|\right].
\end{equation}

We will adapt these general formulas to our case. Since the formulas to be developed here are more relevant for very strong magnetic fields, we also extract from Eq.\,(\ref{FinalQuantization}) the following approximation for the quantized energy levels:
\begin{align}\label{StrongBApproximation}
    E_{n,\ell,s}&=\left(\eta-\frac{1}{2}\right)\Bigg\{m^2\!+\!\left(2n+2\ell+1-\frac{e}{|e|}s\right)|e|B\!-\!\frac{\lambda eBs}{4\hbar}\!+\!\frac{\lambda(2\ell^2+s\ell-2)}{2}\!+\!\frac{\lambda(s\ell+1)}{2\eta-1}\nonumber\\
&\quad+(2n+1+\ell)\Bigg[\lambda (2\ell+s)+(\eta-1)\frac{4\lambda(2n+1-s)}{2\eta-1}+\frac{\lambda s}{2\eta-1}\Bigg]\Bigg\}^{\frac{1}{2}}.
\end{align}
Note that in the absence of gravity, i.e., when $\lambda=0$, this formula reduces to the first identity in Eq.\,(\ref{mTilde}). Let us then start with the consequences of this formula which is valid either (i) for weak gravitational fields and/or (ii) for fermions very close to the center of the star. Since the terms proportional to $\lambda$ in this expression de not contain $B$, it is straightforward to compute the $B$-derivatives in Eq.\,(\ref{GeneralM}). The result for the magnetization remains thus exactly the same as in Eq.\,(\ref{FlatMagnetization}), but with $\tilde{m}_{n,s}^{i2}$ replaced by $E_{n,\ell,s}$ as given by Eq.\,(\ref{StrongBApproximation}):
\begin{equation}\label{CurvedMagnetization}
    \mathscr{M}=\sum_{i=e,\mu,p}\left[\frac{\epsilon_i-\mu^i_fn_i}{B}+\frac{B(2\eta-1)}{4\pi^2}\sum_{s}\sum_{n=0}^{n_{max}}\left(n+\frac{1}{2}-\frac{s}{2}\right)\ln\left|\frac{\mu^i_f+k^i_{f,n,s}}{E^i_{n,\ell,s}}\right|\right].
\end{equation}
We neatly see the general-relativistic correction factored out in the second term of the sum. More important, however, is the fact that the {\it general expression} of the magnetization remains unaltered in this case.

On the other hand, for the case of a {\it non-negligible} gravitational field the quartic term in Eq.\,(\ref{1stOrderFinalfEq}) should be taken into account, leading to the result (\ref{QEpsilonN0}) for $E_{n,\ell,s}$ at the first-order term of the JWKB approximation. As the Landau levels are completely destroyed, we see that even at this order in the approximation the magnetization is completely different from the expression (\ref{CurvedMagnetization}). Actually, even Formulas (\ref{EnergyAndNumberDensity}) and (\ref{mTilde}) do not hold anymore, for such formulas were derived by assuming the linear dependence (\ref{mTilde}) of $\tilde{m}_{n,s}^{i2}$ on the integer $n$. As expression (\ref{QEpsilonN0}) of $E_{n,\ell,s}$ in terms of $n$ displays no such linearity, and since that expression (\ref{QEpsilonN0}) is itself only an approximation for the highly nonlinear full \mbox{expression (\ref{QuantizedEpsilon}),} we conclude that the Landau-levels-based formalism for computing the magnetization $\mathscr{M}$ of neutron stars does not generally apply to the charged plasma in the core of highly dense neutron stars. The condition for the applicability of the formalism is to have the quartic anharmonic oscillator potential term in Eq.\,(\ref{Anharmonic}) negligible, i.e., to have $\lambda\ll |e|B/\hbar$ or to apply the formalism only to fermions close to the vicinity of the star's center. In other words, we conclude that only for a magnetic neutron star of mass $M$, of radius $R$ and with a magnetic field $B$ such that $|e|BR^3c^2\gg GM\hbar$, that the usual formalism generally applies to the charged matter inside the core of the star.

Magnetization in neutron stars affects both the equation of state of the stars' matter and the stars' structure and properties, such as their masses and transport properties. Indeed, both the pressure anisotropy and the softening of the equation of state of the stars matter is believed to be related to magnetization and the distribution of the charged fermions as they occupy the Landau levels. The equation of state determines, in turn, the mass-radius relation of neutron stars (see, e.g., the recent work \cite{EOS2020} and the \mbox{references therein}).  

It is believed that the net magnetization in neutron stars remains negligible until around $10^{16}$\,G, beyond which it starts increasing and becomes oscillatory under the de Haas-van Alphen effect. Since we just saw that the Landau levels become dramatically altered only for non-negligible gravitational fields compared to the magnetic interaction, we conclude that such oscillations will not be much affected in typical stars. For weaker magnetic fields, however, the general-relativistic correction we found in Eq.\,(\ref{CurvedMagnetization}) for the magnetization might leave an observable signature on the equation of state of the star's matter and, hence, on the mass-radius relation of the star as well. More work dedicated specifically to such an interesting investigation is required though.
\subsection{At a Laboratory Level}
Another application of our results is to consider physical systems at the laboratory level. In this case, an interesting system would be the free electrons moving inside a massive metallic sphere put inside a uniform magnetic field. For this case, the gravitational contribution is much weaker than the magnetic interaction, so Eq.\,(\ref{FinalQuantization}) is what we need to apply here. Expanding that expression up to the first order in $\lambda$, we obtain
\begin{align}\label{FinalQuantizationWeakG}
E_{n,\ell,s}&\approx\left(\eta-\frac{1}{2}\right)\Bigg\{m^2c^4+\hbar c^2\left(2n\!+\!2\ell\!+\!1\!-\!\frac{e}{|e|}s\right)|e|B-\frac{\hbar c^2\lambda eBs}{4}+\frac{\hbar^2 c^2\lambda(2\ell^2\!+\!s\ell\!-\!2)}{2}\nonumber\\
&\quad+\frac{\lambda\hbar^2c^2(s\ell+1)}{(2\eta-1)}\frac{\sqrt{1+(2n\!+\!1\!-\!s)\frac{\hbar eB}{m^2c^2}}}{1+\sqrt{1+(2n\!+\!1\!-\!s)\frac{\hbar eB}{m^2c^2}}}+(2n+1+\ell)\hbar^2c^2\lambda\Bigg[\frac{2 m^2c^2}{\hbar eB}+2\ell+s\nonumber\\
&\quad+(\eta-1)\frac{4(m^2c^2+[2n+1-s]\hbar eB)}{\hbar eB(2\eta-1)}+\frac{ s}{2\eta-1}\frac{\sqrt{1+(2n+1-s)\frac{\hbar eB}{m^2c^2}}}{1+\sqrt{1+(2n+1-s)\frac{\hbar eB}{m^2c^2}}}\Bigg]\Bigg\}^{\frac{1}{2}}.
\end{align}
These are the energy levels of the electrons inside the massive sphere. The $\lambda$-terms are the corrections brought to the familiar levels observed in the absence of gravity. This expression is valid for both relativistic and non-relativistic fermions, and it reduces to the familiar relativistic Landau levels in the absence of gravity. This result offers novel ways for exploiting the Landau levels of fermions in the presence of gravity at the \mbox{laboratory level.}

\section{Summary}\label{sec:VI}
We have obtained the Dirac equation for a charged fermion in a general static and spherically symmetric curved spacetime in the presence of a static and uniform magnetic field. We applied the general equation we obtained to the case of fermions in the {\it interior} Schwarzschild solution describing the gravitational field inside a massive sphere of uniform density, and we derived the quantized energy levels of the charged particles. We found that our result reduces to the familiar flat-spacetime relativistic Landau levels of charged fermions inside a uniform magnetic field only when the gravitational interaction is weaker than the magnetic interaction and/or when the charged particles move very closely to the center of the spherical mass. 

We then discussed, based on these results, the consequences on the physics of neutron stars. We found that the magnetization of the core of the latter would dramatically be altered for an extremely high gravitational field compared to the magnetic interaction and/or when not focusing solely on the fermions in the vicinity of the center of the star. We arrived at the conclusion that in the general case, the usual formalism for extracting the magnetization of neutron stars applies only when the latter satisfy the condition $|e|BR^3c^2\gg GM\hbar$. Although we arrived at such a conclusion by relying on a first-order approximation in the gravitational parameter $\lambda$, the generality of our results is not affected. In fact, such a first-order approximation remains valid for any relative strengths of the gravitational and magnetic interactions experienced by the charged particles inside typical neutron stars. Any extra correction to those levels would be obtained simply using time-independent perturbation theory.

In Section \ref{sec:IV}, we outlined a few refinements that might be brought to our model and the prospective outlook on future improvements that will allow us to go beyond the limitations we imposed on our present study. However, even with the simple model dealt with here, we have already glimpsed with new insights into the contribution of gravity in shaping the magnetization of highly compact neutron stars and into novel ways of exploiting the dynamics of charged fermions in the presence of gravity at the laboratory level.

\section*{Acknowledgments}
The authors are grateful to Patrick Labelle for the helpful discussions and comments, and to the anonymous referees for their constructive comments that helped enrich our presentation. This work was supported by the Natural Sciences and Engineering Research Council of Canada (NSERC) Discovery Grant No. RGPIN-2017-05388 and by the Fonds de Recherche du Québec - Nature et Technologies (FRQNT). PS acknowledges support from Bishop's University via the Graduate Entrance Scholarship award.

\appendix
\setcounter{section}{0}
\section{Explicit expressions for \boldmath{$N_k(\kappa,\tau,\ell)$}}\label{sec:App}
Setting $\zeta=[2\Gamma(\tfrac{3}{4})/\Gamma(\tfrac{1}{4})]^2$, where $\Gamma(z)$ is the gamma function and $\varsigma=(\sqrt{2}/\tau)^{\frac{4}{3}}\kappa^2$, the explicit expressions of the nine terms $N_k(\kappa,\tau,\ell)$ for $k=0, ..., 8$ appearing in E.\,(\ref{QuantizedEpsilon}) are given by \cite{JMP(1986),EPJA(2014)}:
\begin{align}
N_0&=1,\qquad N_1=\frac{\varsigma\zeta}{2},\qquad N_2=-\frac{\varsigma^2}{32}\left(1-3\zeta^2\right),\qquad N_3=\frac{\sqrt{\zeta}}{24}\left(6+\frac{\varsigma^3\zeta^2}{8}-12\ell^2\right),\nonumber\\
N_4&=-\frac{\varsigma}{192}\left[2+6\zeta^2-\frac{\varsigma}{64}\left(1-5\zeta^4\right)-12(1+\zeta^2)\ell^2\right],\nonumber\\
N_5&=\frac{\varsigma^2\zeta}{1280}\left(10-\frac{\varsigma^3}{16}-60\ell^2\right),\nonumber\\
N_6&=\frac{1}{192}\Bigg[-\frac{11}{8}-\frac{15\zeta^2}{8}+\frac{5\varsigma^3}{64}(-1+4\zeta^2+\zeta^4)+\frac{7\varsigma^6}{6144}(3\zeta^2+\zeta^6)\nonumber\\
&\quad+\left(25-15\zeta-\frac{\varsigma^3}{16}(5+60\zeta^2-5\varsigma^4)\right)\left(\frac{1}{4}-\ell^2\right)+(10-30\zeta^2)\left(\frac{1}{4}-\ell^2\right)^2\Bigg],\nonumber\\
N_7&=\frac{7\varsigma\zeta}{768}\Bigg[-\frac{39}{4}+\frac{3\zeta^2}{4}+\frac{\varsigma^3}{64}\left(\frac{7}{2}-\frac{10\zeta^2}{3}-\frac{\zeta^4}{2}\right)-\frac{\varsigma^6\zeta^2}{1536}\left(\frac{4}{5}+\frac{\zeta^4}{7}\right)\nonumber\\
&\quad+\left(28-6\zeta^4-\frac{\varsigma^3}{32}\left(7+20\zeta^2-\zeta^4\right)\right)\left(\ell^2-\frac{1}{4}\right)-\left(8-12\zeta^2\right)\left(\ell^2-\frac{1}{4}\right)^2\Bigg],\nonumber\\
N_8&=\frac{9\varsigma^2}{4096}\Bigg[\frac{23}{24}+\frac{95\zeta^2}{2}-\frac{9\zeta^4}{8}+\frac{\varsigma^3}{64}\left(\frac{1}{9}-9\zeta^2+3\zeta^4\right)+\frac{\varsigma^6}{4096}\left(\frac{1}{63}+\frac{9\zeta^4}{5}\right)\nonumber\\
&\quad-\left(\frac{31}{3}+96\zeta^2-9\zeta^4-\frac{\varsigma^3}{64}\left(\frac{4}{3}+\frac{192}{5}\zeta^2+36\zeta^4\right)\right)\left(\ell^2-\frac{1}{4}\right)\nonumber\\
&\quad+\left(\frac{14}{3}+16\zeta^2-18\zeta^4\right)\left(\ell^2-\frac{1}{4}\right)^2\Bigg].
\end{align}


\end{document}